# The psychometric house-of-mirrors: the effect of measurement distortions on agent-based models' predictions

Dino Carpentras & Michael Quayle

# The psychometric house-of-mirrors: the effect of measurement distortions on agent-based models' predictions

Dino Carpentras [a,b] and Michael Quayle [a,c,d]

[a]Social Dynamics Lab, Department of Psychology, Centre for Social Issues Research, University of Limerick, Limerick, Ireland; [b]MACSI (Mathematics Applications Consortium for Science and Industry) Department of Mathematics and Statistics, University of Limerick, Limerick, Ireland; [c]The SFI Research Centre for Software, University of Limerick, Castletroy, Limerick, Ireland; [d]Lero, Department of Psychology, School of Applied Human Sciences, University of KwaZulu-Natal, Scottsville, South Africa

**ABSTRACT**

Agent-based models (ABMs) often rely on psychometric constructs such as 'opinions', 'stubbornness', 'happiness', etc. The measurement process for these constructs is quite different from the one used in physics as there is no standardized unit of measurement for opinion or happiness. Consequently, measurements are usually affected by 'psychometric distortions,' which can substantially impact models' predictions. Even if distortions are well known in psychometrics, their existence and nature is obscure to many researchers outside this field. In this paper, we introduce distortions to the ABM community. Initially, we show where distortions come from and how to observe them in real-world data. We then show how they can strongly impact predictions, qualitative comparison with data and the problem they pose for validation of models. We conclude our analysis by discussing how researchers may mitigate this problem and highlight possible future modelling trends that will address this problem.

## Introduction

One of the key features of agent-based models (ABMs) is that they may be used for addressing complex social, phenomena (Abar et al., 2017; Srbljinović & Škunca, 2003). For example, vaccine hesitancy and climate change are both pressing social problems which depend on collective action for their resolution (Johnson et al., 2020; World Health Organization, n.d.). Thus, understanding how people interact, coordinate opinions and behavior, and how to facilitate this process is crucial for the future of our society.

ABMs are designed to explicitly model these complex interactions (Cioffi-Revilla, 2014) to better understand the dynamic process of our society. For example, models of social influence (Castellano et al., 2009; Flache et al., 2017) can be used to identify which countries are more at risk of developing strongly anti-vaccine opinions and behavior in the coming years (Carpentras et al., 2022). This could then be used to produce tailored interventions.

An important issue with ABMs is that they often rely on psychological constructs, such as opinion, stubbornness, or happiness (Deffuant et al., 2000; Flache et al., 2017; Han et al., 2019; Rojas & López, 2014). When ABMs are used to make conclusions about real-world phenomena, it is usually necessary to link the quantities represented in the model to real-world measurements, either to seed the model or to assess its results (Duggins, 2014; Hassan et al., 2008; Innes, 2014; Jia et al., 2015; Valori et al., 2012). However, while measurements in physics are usually based on precise units of measurements, there is no such a thing as a unit of opinion or wellbeing (Kranz et al., 2006a; Stevens, 1951).[1] This means that measurements of these constructs, will be possible, but affected by 'psychometric distortions' (Kranz et al., 2006a; Wright, 1999).

These measurement issues have important consequences for modelling. For example, many studies have already explored how distortions may affect statistical models, especially parametric linear models which have assumptions frequently violated by psychometric measures (Baggaley & Hull, 1983; Feir & Toothaker, 1974; Glass et al., 1972; Lantz, 2013). Furthermore, their effect is so strong that using different ordinal scales for the same measurements can even reverse the predictions of a model (Schroeder & Yitzhaki, 2017). Although there are some papers exploring the complex issues of linking ABMs to empirical data (e.g. Boero & Squazzoni, 2005; Bruch & Atwell, 2015), or discussing the problems posed by number of levels in the measurement scale, the granularity of the time-scale or the problem of incomplete datasets (Hassan et al., 2008; Lee et al., 2015; Troitzsch, 2021) the more specific issue of how psychometric distortions can affect ABM results has not yet received direct attention. This is an increasingly important issue, as it not uncommon for people to compare simulation results to real data or directly 'inject' real data into ABM's to provide realistic starting conditions for models. Indeed, several methodological papers treat this as a straightforward procedure. For example, Bruch and Atwell (2015) argue that 'it is relatively straightforward to assign agents characteristics from these [survey] data' (p. 194). Others go on to directly initialize agents with self-report psychometric data derived from participant surveys (e.g. Carpentras et al., 2022; Pakravan & MacCarty, 2021; Tumasjan & Beutel, 2019; Valori et al., 2012). Similarly, Hassan et al. (2008) 'provide guidelines for data injection' into ABMs, and recommend that 'data values should be used to define the initial population of agents, with their initial values' (p. 8).

While we agree in principle with the importance and value of linking ABMs to empirical data, we note that these calls are usually made without consideration for the compatibility between psychometric data and how this might affect ABMs.

There are two reasons that this issue needs immediate consideration. First, while some computational modelers are aware of this general problem due to their specialized training (e.g. in psychometrics or mathematics), it is also true that the field of agent-based models and social simulation is radically interdisciplinary, and many people designing models do not have specialist training in survey methods, psychometric measurement, or how these interfaces with mathematical models. Second, there are some issues in using psychometric data that are specific to agent-based modelling (Carpentras & Quayle, 2022; Carpentras et al., 2020). As we will discuss, distortions pose particular problems for (a) the practice of validating a model by comparing it against real-world data, especially when this is performed on multiple scales at the same time (Chattoe-Brown, 2014; Duggins, 2014; Wright & Sengupta, 2013), and (b) seeding models with real-world data, both of which are recommended by seminal papers on empirically grounding ABMs (e.g. Bruch and Atwell, 2015; Boero & Squazzoni, 2005; Hassan et al., 2008).

Since the ABM community is composed by people with extremely different backgrounds, we do not presume that readers have background knowledge of psychometrics, and we will start by introducing ordinal scales and how they can produce distortions in the data. Readers who are already knowledgeable about psychometric distortions may safely skip this. Then, we will show how these distortions may have an important impact on the model's quantitative and even qualitative predictions; thus, also affecting the processes of validation and calibration. In the conclusion, we will discuss the possible effects of this problem and how it can be managed.

## Measurement theory

### Introduction on ordinal scales

To understand distortions, we need to first introduce *ratio scales* and *ordinal* scales.[2] While most researchers with a background in the social sciences have already heard of these terms, they are sometimes used to support some rather counterintuitive claims. For example, it is often said that ordinal scales do not support arithmetic operation, such as addition and subtraction (Merbitz et al., 1989; Ricotta & Avena, 2006; Wu & Leung, 2017). Or, similarly, that the difference between two ordinal scores, for example, score 6 and score 5, does not exist. In this section, we will introduce ordinal scales and, hopefully, we will be able to clarify how they work, why they are often associated with such claims and how they can produce distortions.

To explain ratio scales, let us start with an example. Let us pretend that we want to measure the height of some people using pencils of same length. Thus, we stick them one over the other, we number them, and then we start measuring people in 'pencils.' Meaning that the first person's height would be 11 pencils, the second person would be 9 pencils tall, etc.

Now, let us imagine trying to perform the same measurement of height without any tool. In this case, we can still measure people by ordering them from the shortest to the tallest (or vice versa). Even if this will not tell us the exact height of everyone, it still allows us to make claims such as 'person 6 is taller than person 5' or 'person 1 is the shortest.'

Now, let us put aside this example and let us try to summarize what it taught us. In the first case, we used the pencils as a unit of measurement. Scales of this type are called 'ratio scales' and are similar to most of the measurements we use in everyday life. Let us stress that expressions like '3 meters' or '5 pencils,' where intended as measurements, can be represented as a multiplication. Indeed, '3 meters' means '3 times the length of a meter.' As we will see, this is not the case with ordinal scales.

The other type of scale we faced in our example is a little less common and sometimes people will not even consider it a proper measurement. For example, saying that someone is the second richest person in the world tells us that this person is extremely rich, but not how much money they have. Since these scales preserve order and nothing more, they are usually referred as 'ordinal scales.'

Unfortunately, we are so used to ratio scales that ordinal scales are often really hard to imagine. Thus, people often visualize them as rulers with uneven spacing between their ticks, or, similarly, as an elastic ruler (Forrest & Andersen, 1986; Gardner, 1975). Indeed, with such a ruler, we could say that an object of length 10 is longer than an object of length 9, but not how much longer. Indeed, the distance between mark 10 and mark 9 of such a ruler may be very small, while the distance between marks 9 and 8 may be way bigger. This is also visualized in Figure 1 where we see that the differences in height between people are not equally spaced. Also, notice the relationship between measurements on this scale cannot be interpreted as multiplication. Indeed, being as tall as person 9 does not mean that our height is 9 times the height of person 1.

A trick to avoid confusion in this case is to consider each value as a label. For example, if we use a ruler (either evenly or unevenly spaced) for making a measurement, we will indicate the position of mark $i$ as $m_i$. This makes clear that the difference between, say mark 9 and mark 5, is not calculated directly as 9–5 but as:

$$d(m_9, m_5) = |m_9 - m_5| \tag{1}$$

In the special case in which our scale is ratio, we know that it may be interpreted as a multiplication of a unit. Meaning that mark $m_i$ would be just $i$ times the length of $m_1$:

$$m_i = im_1 \tag{2}$$

This formula means that a distance between ticks may be simply calculated as the subtraction of their indices:

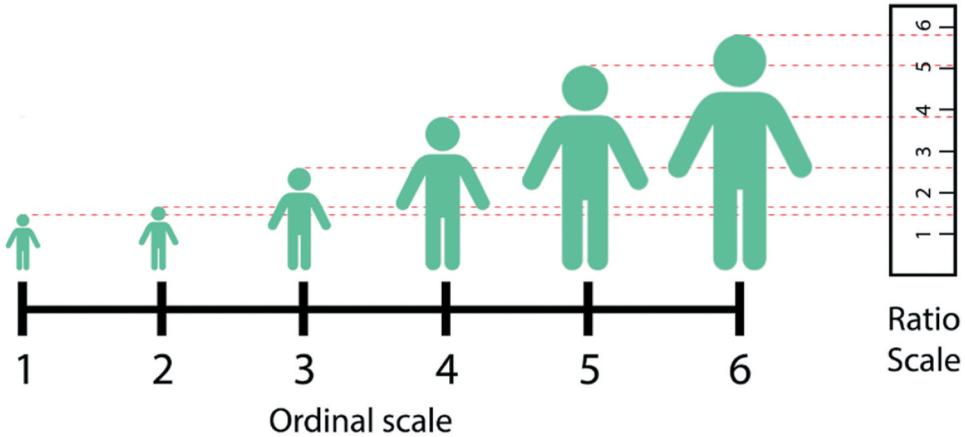

**Figure 1.** Representation of measurement of people's height in an ordinal and a ratio scale.

$$(m_i, m_j) = |i - j| m_1 \qquad (3)$$

Thus, we can immediately calculate the difference between a 9-m object and a 5-m one as (9–5). However, this is not the case with ordinal scales. Indeed, it does not make any sense to say that the difference between person 9 and person 5 is 4 times the height of person 1. This is the reason why we may hear claims as 'ratio scales support arithmetic operations while ordinal scales do not.'

### *Theory of distortions*

Up to now we analyzed ratio and ordinal scales by themselves. However, to observe distortions, we need to analyze what happens when we compare two different scales. To keep our previous example, let us suppose that also a friend had to measure the height of people in another group. However, for the first case, instead of pencils, she measured people using erasers.

As we compare our measurements, we notice that a 10-pencils tall person is 20-erasers tall. Furthermore, a 12-pencils person is also 24-erasers tall, etc. Indeed, we do not need to compare each person just to notice that 1 pencil is equal 2 erasers. Thus, if a person is $i$-pencils tall she would also be $2i$-erasers tall. This relationship between ratio scale is also visually represented in Figure 2 where we can see the comparison of two ratio rulers together with a graph representing their linear relationship. Indeed, the relationship between ratio scales can be written as:

$$M_i = \alpha m_i \qquad (4)$$

Where $\alpha$ is a scalar value, $m_i$ are the positions of the marks in one scale and $M_i$ on the second scale.

However, the situation become more complex when it gets to the ordinal scales. In this case, we may notice, that the shortest person in our group is as tall as person 3 in our friend's group. Furthermore, our person 2 is equivalent to person 5 of the other group. However, this does not tell us almost anything on how to convert the height of person 3.

Indeed, while in ratio scales we can just compare the unit, in ordinal scales we need to compare all the levels one by one. The only information that is preserved from one scale to the other is the order. Meaning that if person a is taller than person b, then the score of a would be bigger (or equal)[3] than the score of b in every ordinal scale. In formula we write:

$$M_i = h(m_i) \qquad (5)$$

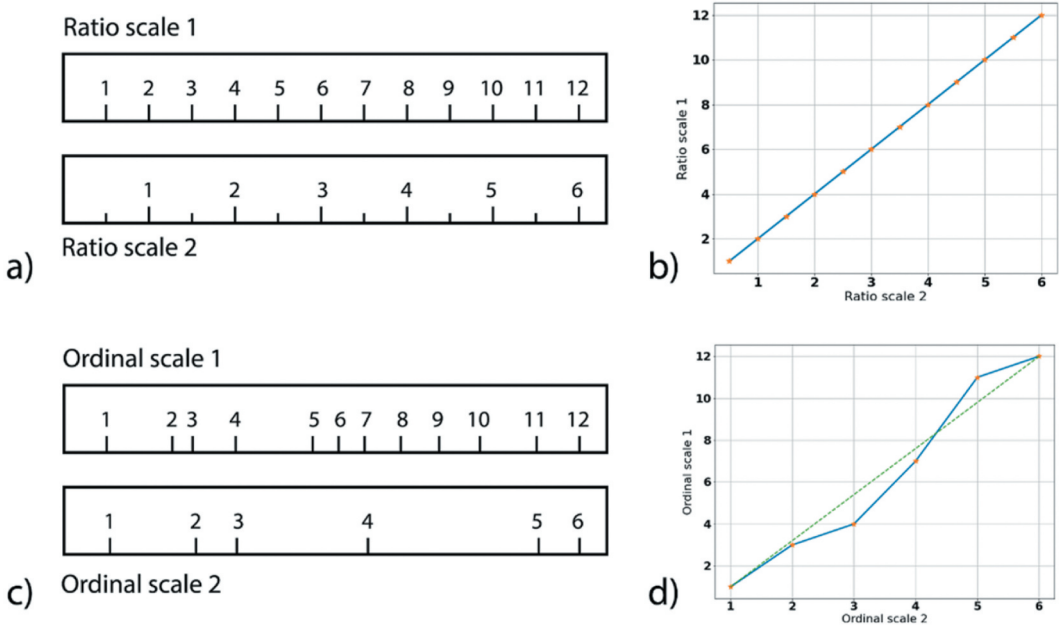

**Figure 2.** Comparison between two scales a) and c) representation of ratio and ordinal scales as rulers b) and d) equivalence between marks on one ruler and the marks in the other ruler.

Where $h$ is a monotonic function which, in general may be non-linear (Krantz et al., 2006a; Schroeder & Yitzhaki, 2017). Here 'monotonic' is just the mathematical term for 'order-preserving function'.

Figure 2(c-d) shows the relationship between two ordinal scales both in the form of rulers and as a graph. For the rulers, we may notice how knowing the relationship of the first 3 marks does not tell us how to convert mark 4. In the graph, this is visually represented by the non-linear relationship between the two scales.

## Ordinal scales and distortions in psychometrics

### From psychometric measurement to distortions

Up to now we discussed the theory of ordinal scales using a very unrealistic example. In everyday measurements, why should we care about ordinal scales when we can directly use ratio ones? The problem is that, for many psychological constructs no ratio scale exists[4] (Krantz et al., 2006a; Stevens, 1946, 1951).

Indeed, when we are measuring distance, we can take two rods of equal length, stick them one after the other and obtain a rod twice as long as the original ones. Similarly, we can take n rocks of 1 kg, put them in a bag, and obtain an overall weight of n kg. In measurement theory this operation is called 'concatenation,' (Krantz et al., 2006a) and it is used to produce ratio scales from whatever initial unit. Indeed, to make a measurement we just concatenate our unit n times to claim that our object is n units long (or heavy).

In psychometric measurement, this operation is rarely possible. Indeed, if we take two people having the same level of happiness, there is no way we can concatenate them to obtain a new person twice as happy as the original ones. Psychometrics constructs in general cannot be concatenated making it impossible to produce units for them. This is the reason why in psychometrics, constructs are usually measured in ordinal scales (Krantz et al., 2006a; Stevens, 1946, 1951).

Another way to understand why psychometrics measurements are ordinal and not ratio is to observe, practically, how measurements are performed. One of the most common ways to measure psychometric constructs is via self-rating (Nunnally, 1978). Thus, if we want to measure how much a person likes a certain movie, we can ask her to select for the following item the choice that better represents her opinion.

'This is a good movie.'

[] Strongly disagree [] Somewhat disagree [] Neutral [] Somewhat agree [] Strongly agree"

Starting from this set of response-options we can definitely tell that 'Strongly agree' is more positive than 'Somewhat agree,' however, we do not know how much more positive it is. Thus, what we obtained is a scale which tells us order and nothing more; i.e. an ordinal scale.[5]

A very common (and potentially misleading) approach is to code these answers into numbers, such as 'Strongly disagree' = 1, 'Somewhat disagree' = 2, etc. Even if standard literature informs us on the fact that these are ordinal scales and so we should not use arithmetic operations (Merbitz et al., 1989; Ricotta & Avena, 2006; Wu & Leung, 2017), it is still tempting to claim that the difference between 'Strongly disagree' and 'Somewhat disagree' is 1. Thus, we can easily think that we are dealing with a ratio scale and completely forget about distortions. To avoid confusion, instead of numbers we can use labels as we did before, so that 'Strongly disagree' = $m_1$, 'Somewhat disagree' = $m_2$, etc.

Another problem regards to how we can build different scales measuring the same construct. This situation is present even in physics, where everyone can come up with her own measurement of distance. Indeed, even in the modern world different countries use different measurements of distance and weight (Halliday et al., 2013). However, as we know, if measurements are ratio (i.e. based on a unit), we can convert them with a simple linear relationship.

As we may expect, this problem is similar in psychometrics, where everyone can produce a different (ordinal) measurement of the same construct. For example, to measure trust in vaccination, we may ask a person to rate their feeling towards vaccines on a scale 0 to 10 (European Social Survey, n.d.). Alternatively, instead of numbers, we may use the possible answers: 'Trust vaccines,' 'Neutral,' 'Do not trust vaccines.' We could also add some other possible answers, such as 'Strongly trust' and 'Weakly trust,' (Wellcome Trust, n.d.). Otherwise, we can ask multiple similar questions, such as 'Do you think vaccines are important for children?' 'Do you think vaccines are safe?' etc. and then combine their answers to obtain a single score (Everett, 2013).

Of course, a different choice of questions or response options will result in a different ordinal scale. This does not mean that one scale would be better or worse than the other. Indeed, an ordinal scale is good as long as it preserves the right order (Krantz et al., 2006a). However, as we will see in the next section, different scales will present different distortions.

### *Distortions in real data*

Up to now we discussed about distortions in data from fictious scenarios. Here we want to observe them in real-world data. To do this, we use data from the (Wellcome Trust, 2018; Wellcome Trust, n.d.), a dataset on more than 140,000 people. According to the manual, five questions were designed to measure the construct 'trust in science' (Wellcome Trust, n.d.). This means that their answers can be averaged together to produce an overall (ordinal) score of how much each person trusts science.

To observe psychometric distortions, we need to compare two different ordinal scales, which we will construct from different combinations of the items selected by the survey designers. We can think of this thought experiment as exploring two parallel universes where the survey designers used 3 questions instead of 5. The first scale is obtained by combining items 1 to 3, and the other by combining items 3 to 5 (thus item 3 contributes to both scores).

By looking at Figure 3(a) we can see that the relationship between the two scores is clearly a nonlinear one. For example, people with a score of 4 on the first scale, will on average obtain a similar score on the second scale. However, scoring 10 on scale 1 is equivalent to scoring 7 on scale 2.

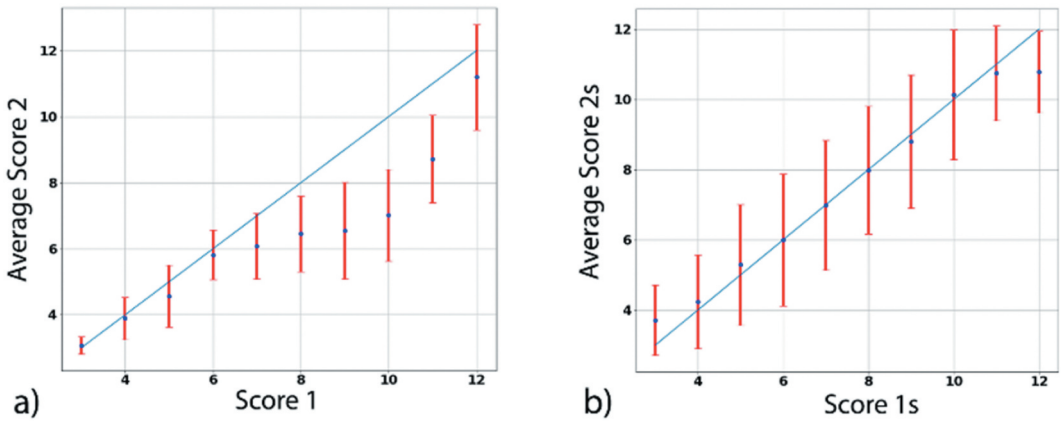

**Figure 3.** (a) relationship between the scores on two experimental scales (each consisting of different items on trust in science) represented as the red error bars. The blue line represents the ideal linear relationship, while the deviations from this line of actual data points show the presence of psychometric distortions. (b) simulated data for two scales in presence of strong noise but no distortions.

Notice that Figure 3(a) also shows an error bar which was not discussed in our previous analysis. This is due to measurement noise and it is not what we want to focus on in this article. However, it is good to show here that noise is a different phenomenon from psychometric distortions. Indeed, noise represents a mistake in the ordering, thus degrading the quality of ordinal scales. Instead, distortions are just due to the lack of a unit of measurement and do not represent a lower quality of measurement (Krantz et al., 2006a, 2006b, 2006c). To better show this difference, in Figure 3(b) we simulated the case of 2 scales affected by noise but not distortion. As we can see, the error bar is present, but the relationship between the two is linear except for the extremes which exhibit ceiling and floor effects.

To better visualize distortions, we can think a house of mirrors. Here, every curved surface produces a new image of us which still preserves order but not proportion. Indeed, in each image, our head will be above our neck, but it will be relatively longer or smaller in different mirrors.

We could use these reflections then, to look at our body and list the parts of the human body from top to bottom. As this task only needs ordering, we could use any mirror for this task and obtain the same result. However, if we try to use the reflections to measure proportions, every mirror will provide us with a different result. Thus, if we forget about their 'ordinal nature,' we may think that our head is changing size as we move through the hall of mirrors.

This means that ordinal scales are still good at measuring what they intend to measure (i.e. ordering of levels in a specific construct). However, when using them in a model, we need to remember their ordinal nature, and so the possible distortions they may introduce.

## Seeding models with real-world data

### The effect of distortions on the distribution

In the previous sections, we discussed where distortions come from and showed that they can affect real-world data. Thus, if we load real-world data into ABMs, their predictions may also be affected by distortions. Similarly, comparing a model's output to real-world data (either a distribution or a single value such as the mean) may produce different results depending on the scale we are comparing to. However, we have still not clarified how distortions impact the collected data and, consequently, the model's predictions.

To make this effect more explicit, we will focus on the distribution of the data. Specifically, we suppose that three researchers (let us call them Alice, Bob and Claire) independently collected 3 measurements of the same psychometric construct (e.g. political attitudes) on the same group of people. However, as each one created a scale independently from the others, they end up with three different ordinal scales.

In previous sections, we called the $h$ the function that maps data from one scale into the other. Furthermore, we said that $h$ is order-preserving and, in general, non-linear. Thus, we call $h_B$ the function that maps Alice's data to Bob's one and $h_C$ the one that maps Alice's data to Claire's one.

To analyze the effect of a scale change, we simulated that every researcher produced an 11-points score (from 0 to 10). The simulated population consisted of 100,000 participants whose opinions are uniformly distributed in Alice's data. We can observe this from the fact that the histogram in Alice's data is almost flat (Figure 4).

To obtain Bob's and Claire's data we defined the $h$ functions. As we can see, from Figure 4, $h_C$ has an s shape, while $h_B$ has an inverse s shape. By applying these transformations, we obtain the two distributions shown in Figure 4. As we can see, even if the three (simulated) datasets are measuring the same construct on the same people, their distributions are qualitatively different. Indeed, while Alice's distribution is uniform, Bob's one is monomodal and Claire's one is bimodal. This shows us that changing ordinal scale (even if we are measuring the same construct on the same people) may strongly impact the data distribution.

To make sense of this effect we can observe the shape of the functions $h$. Indeed, we can see that $h_B$ has a plateau between levels 4 and 6. This means that all the people which scored between 4 and 6 on Alice's scale will score 5 on Bob's one. This explains why we observe the central peak in Bob's distribution.

This effect is also visible in real-world data. Indeed, in previous sections, we obtained two scales of trust in science from the Wellcome Global Monitor. As expected from ordinal scales, by analyzing the data we showed that the function mapping data from one scale to the other was non-linear. In Figure 5 we can observe how this non-linearity is reflected in data distribution. Indeed, the two histograms shows us the data distributions when measuring trust in science in the two scales.

The distribution of scale 2 presents a strong peak (more than 3 times all the other bars), while nothing similar appears in the other distribution. Furthermore, the distribution of scale 1 is quite symmetric while, the distribution of scale 2 is more skewed. In the next sections, we will discuss how this can impact processes such as model's predictions, comparisons with real-world data and calibration.

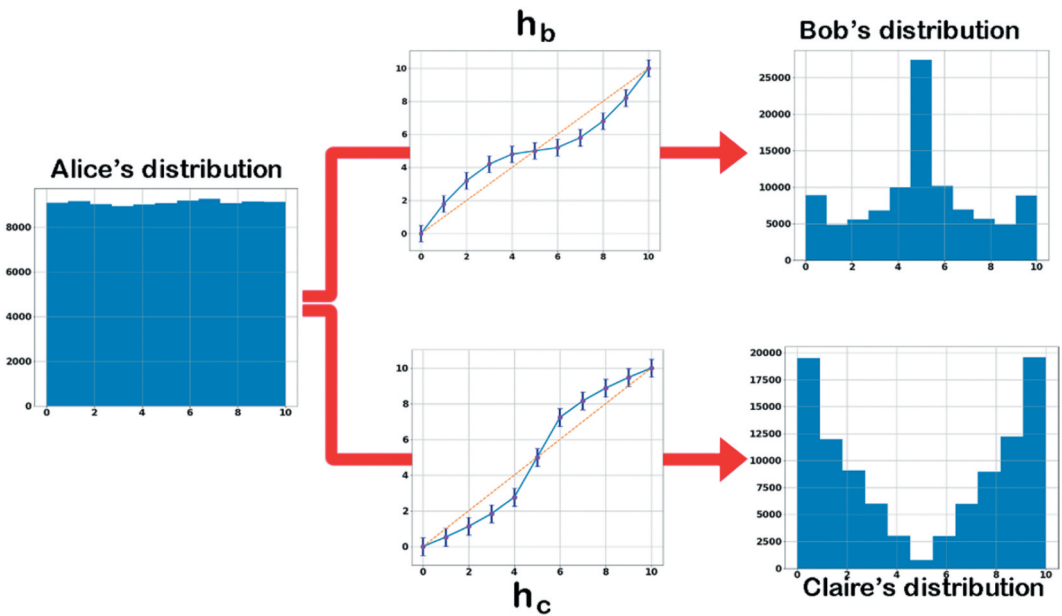

**Figure 4.** Distributions obtained from three different ordinal scales measuring the same construct. The function h maps the values from one distribution to the other.

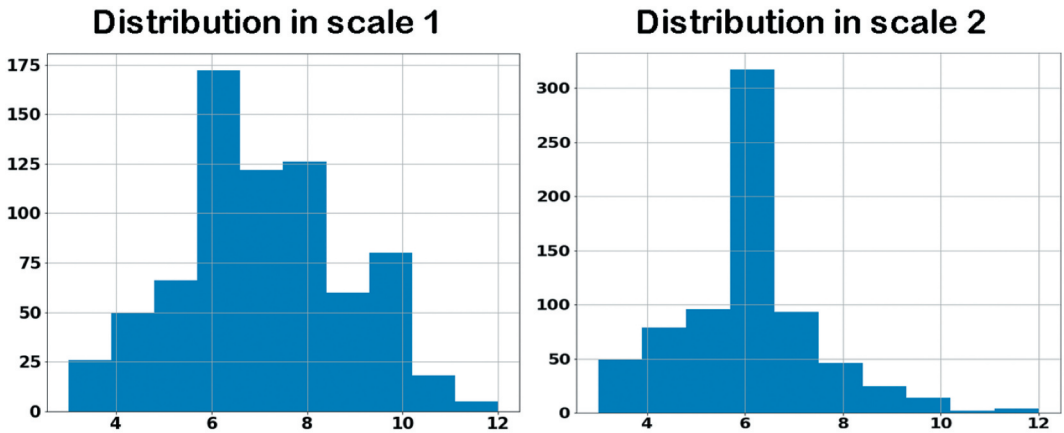

**Figure 5.** Distribution of trust in science for the two ordinal scales.

### *Test on the Deffuant model*

Since distortions have a strong effect on data distribution, we may expect that they will also impact the model's dynamics. In this section, we will study how this can impact the mean output of the Deffuant model (Deffuant et al., 2000). We chose this model as it has been intensively studied, it has few free parameters and a very straightforward dynamic. Furthermore, the Deffuant model has the convenient property that the average of the output distribution does not depend on the choice of the parameters, nor on the running time. To show that the issues we observe are not limited to the Deffuant model, in the next sections we will also verify this impact on three different versions of the Hegselmann-Krause model (Hegselmann & Krause, 2005) and on the Duggins model (Duggins, 2014).

In the Deffuant model, each agent has an opinion $x$, represented as a real number. When this agent interacts with another agent, she will move towards the average opinion of both. As mentioned, the Deffuant models has a few simple parameters. The first one, usually referred as μ, regulates mostly the convergence speed (Deffuant et al., 2000). The second parameter is usually referred as ε and represent the maximum opinion distance at which two agents can interact. Meaning that if their difference of opinion is bigger than ε, they will not interact and cannot change each-other's opinion. We tested the simulations on a uniform distribution and different values of μ (i.e. 0.2 and 0.5) and ε (i.e. 2, 4 and 5), but, as expected, they did not affect the final average.

We initially tested the effect of distortions on simulated data. For producing scale 1, we simulated 1,000 agents with a uniform starting distribution. We obtained the data from scale 2 by applying a non-linear transformation to scale 1 to introduce the types of distortions discussed above. For simplicity, we choose transformations of the type $x^n$ and $x^{\frac{1}{n}}$. The error is calculated as the difference of the mean final opinion from scale 1 and the mean final opinion from scale 2.

Results are shown in Table 1. Already, for distortions such as $x^2$, we have that model outputs may differ up to 17% of the scale. We also see that if we introduce massive distortions, such as $x^{100}$, we could expect to have a prediction error as big as the entire scale.

We also ran the Deffuant model seeded with real-world data from the Wellcome Global Monitor, using the two scales described above on trust in science, and repeating the analysis for every country in the dataset. For each country, we calculated the mean final opinion for data in scale 1 and scale 2. We found that predictions between the two scales could have differences as big as ±26%. Furthermore, for some countries, we found that by using one scale we predicted that people will be overall positive towards science, while using the second scale we predicted the opposite. This is similar to what has been found by Schroeder and Yitzhaki (2017) for linear regression models where 'monotonic increasing transformations can in fact reverse the conclusion reached.'

Table 1. Predictions difference for different scale transformations in simulated data for the Deffuant model.

| Distortion's magnitude (n) | 1 | 2 | 5 | 10 | 100 |
| --- | --- | --- | --- | --- | --- |
| Error | 0% | 17% | 34% | 41% | 50% |

### Test on the Hegselmann-Krause models

To show that this situation is not limited to the Deffuant model, we tested the Hegselmann-Krause (HK) model (Hegselmann & Krause, 2002), together with two of its variations (Hegselmann & Krause, 2005), in the same way. This model is similar to the Deffuant model, except for the fact that agents do not interact in pairs but in groups. Indeed, at each time step, one agent is selected together with all the agents with 'similar' opinions. Where 'similar' means that the distance is less than the bounded confidence parameter ε. During this interaction, all the interacting agents update their opinion with the arithmetic mean opinion of the group. The two alternatives of the HK model are different only as they use the geometric and harmonic mean.

We run our simulations with 1,000 agents and ε = 0.3. We run the model for 100 steps, which is enough to ensure convergence of the model (Hegselmann & Krause, 2005).

Table 2 shows the error level for the different configurations. In this case, we used two different columns (one for distortions of the type $x^n$ and another one for distortions of the type $x^{\frac{1}{n}}$) as the two different types of distortions produced different errors even for the same n.

The two variations of the HK model seem a little more robust to distortions of the type $x^n$ as, even for distortions of the type $x^{100}$ we never exceeded an error of 46%. However, this does not mean that it is unaffected. Indeed, even for distortions as small as n = 2 we already have an uncertainty of ± 17% on the full scale. Furthermore, this robustness to distortions of the type $x^n$ is someway compensated by the stronger error produced by distortions of the type $x^{\frac{1}{n}}$ which can be as big as 55% of the total scale.

Also in this case, we repeated the analysis with the Wellcome Global Monitor data finding that the difference of average may be as big as ± 30% for the harmonic mean model (see, Figure 6). Figure 7 shows another interesting result which appears from the convergence patterns of the model when seeding it with the distributions displayed in Figure 5. In both cases, the model produces a mean peak around 6. Furthermore, both final patterns have a second minor peak. The interesting point is that in one case this secondary peak is located in 3, while in the other case is in 10. Therefore, the pattern shows in one case a tail towards more positive values and, in the other case, a tail towards more negative values. This already shows us how distortions may produce also visually qualitative differences in model's output. We will discuss this in more detail in the next section.

### Test on the Duggins model

In the previous sections, we choose the Deffuant and HK model as they are both well-know and simple. Thus, it is easy for the reader to see that the observed effect is only due to distortions. However, these models can be considered too simple, and we may wonder if distortions may affect also more modern and complex models. Because of that, we choose the Duggins model of opinion dynamics (Duggins, 2014). This model is not based on the bounded confidence principle, has shown complex dynamic patterns, and has also been applied to real-world data.

Table 2. Output difference for different scale transformations in simulated data for different versions of the HK model.

| Distortion's magnitude (n) | | 1 | 2 | 5 | 10 | 100 |
| --- | --- | --- | --- | --- | --- | --- |
| Arithmetic mean | $x^n$ | 0% | 17% | 34% | 41% | 50% |
| | $x^{\frac{1}{n}}$ | 0% | 17% | 34% | 42% | 50% |
| Harmonic mean | $x^n$ | 0% | 17% | 30% | 37% | 43% |
| | $x^{\frac{1}{n}}$ | 0% | 18% | 38% | 46% | 55% |
| Geometric mean | $x^n$ | 0% | 17% | 33% | 39% | 46% |
| | $x^{\frac{1}{n}}$ | 0% | 18% | 36% | 44% | 53% |

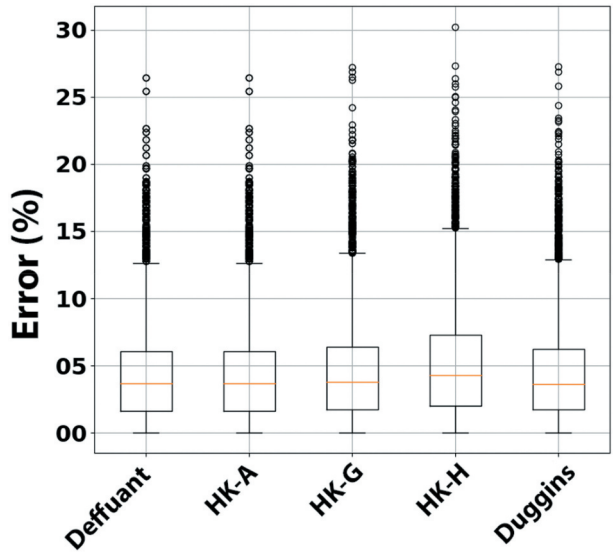

Figure 6. Boxplot of the output error for different models while using the two scales from the Wellcome Global Monitor for different countries (i.e. every data point represents a different country). We can see how each model for these data can make predictions with errors above ±25%.

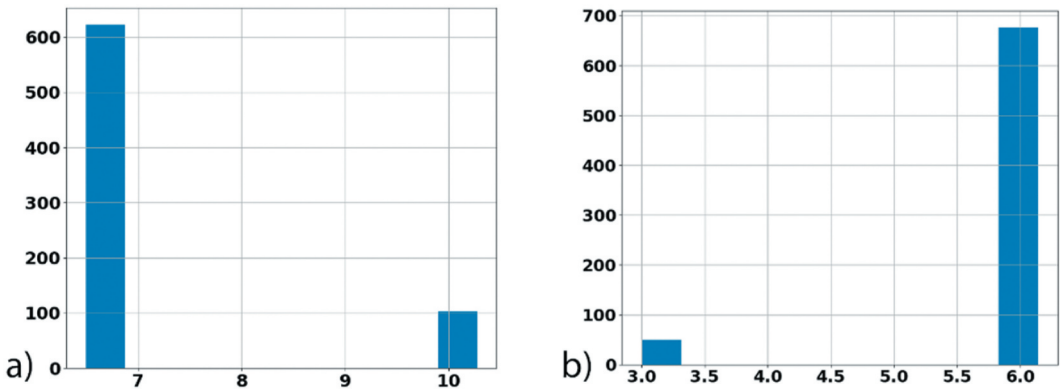

Figure 7. Qualitative difference of model's output when seeding the classic HK model with real-world data having different distortions.

Duggins model is a model of opinion dynamics which also includes the effect of conformity. It does not rely on a threshold as the bounded confidence models but it uses parameters such as conformity (parameter $c$), susceptibility ($s$) and tolerance to dissimilar opinions ($t$). Furthermore, these parameters can be different for every agent (while in the previous models all agents shared the same threshold).[6]

In our simulations, all parameters have been chosen from a random distribution similar to how performed in the original article. This means that every agent may end up with a different set of parameters. These parameters do not change during the model's dynamics, while agents' opinion can. We used 1,000 agents, $s$ and $t$ have been randomized between 0 and 1, while $c$ has been randomized between −1 and +1.

Also for the Duggins model we repeated the simulations for different distortions levels finding results comparable with the previous models (see, Table 3). When we then applied this model to real-world data we found predictions errors as big as $\pm 28\%$ (see, Figure 6). This shows us that even more complex and realistic models can be as well strongly affected by distortions.

Table 3. Predictions difference for different scale transformations in simulated data for the Duggins model.

| Distortion's magnitude (n) | 1 | 2 | 5 | 10 | 100 |
| --- | --- | --- | --- | --- | --- |
| $x^n$ | 0% | 20% | 36% | 43% | 50% |
| $x^{\frac{1}{n}}$ | 0% | 18% | 34% | 41% | 49% |

## Comparing the model's output to real-world data

In the previous section, we focused mostly on how seeding a model with real-world (distorted) data can produce prediction's errors and uncertainty. Here instead, we want to focus on how distortions impact the practice of qualitatively comparing model's outputs to real-world data.

For this analysis, we suppose that the dynamic of opinions in the real world follows exactly a model M. Here, for simplicity, we choose the Deffuant model, but the same argument can be carried out with any other model producing outputs which can be compared to psychometric measurements.

In Figure 8(a) we see the opinion distribution at time 1 for 1,000 agents/people. Figure 8(b) represents the opinion after a time $\Delta T$ corresponding in the simulations to 1,000 iterations (we chose $\mu = 0.5$ and $\varepsilon$ is equal the scale range, meaning that everyone can interact with everyone else).

Notice also that, without mentioning it, we had to assume a scale of opinion for performing this operation. However, this scale (which we can call the 'real scale' or the 'ideal scale') may be different to the scale in which these opinions are practically measured. This means that while there may be a scale S in which the model M perfectly represents the dynamics of opinions, in general this will not be the same scale used for collecting the data.

For this example, we used the same distortion $h_c$ we used for Figure 4. In Figure 8(c-d) we represented the data at time 1 and 2 in the measurement scale. Therefore, they represent how the collected data from time 1 and 2 will actually look like.

The first striking result is that the observed data have a pattern which resembles the irregular patterns of real-world data, despite following perfectly the Deffuant model. If we compared the patterns of the Deffuant model against these data – as done or suggested in the literature (Chattoe-Brown, 2014; Duggins, 2014; Flache et al., 2017) – we would (erroneously) conclude that the Deffuant model cannot reproduce their dynamics.

We can confirm this also by taking the collected data at time one (i.e Figure 8(c)) and seeding them in the Deffuant model. By using exactly the same parameters as before (i.e. 1,000 iterations, 1,000 agents, $\mu = 0.5$ and maximal $\varepsilon$) we obtain the result shown in Figure 8(e) which is qualitatively very different to Figure 8(d).

## The falsifiability problem

Although many ABMs are designed as purely theoretical 'toy' models, there are many applications where we would like to use ABMs to model real-world processes. As soon as we want to use ABMs for making predictions in the real world, it becomes important to distinguish between 'good' and 'poor' models, specifically, rating the model's quality based on how precise their predictions are.

In the world of physics this can be done in a fairly easy way. Indeed, we can take data from the real world at time $T_1$, use them to seed the model, make our prediction and then compare it with the data of the real-world at time $T_2$. The more the prediction resembles the real data, the better the model.

Even if data injection into ABM models has been suggested and applied (Chattoe-Brown, 2014; Duggins, 2014; Flache et al., 2017; Hassan et al., 2008) this process can be much more complex when it involves psychometric data. Indeed, as we saw before, the prediction does not depend only on the model but also on the chosen scale. Indeed, a change of scale may strongly alter each prediction.

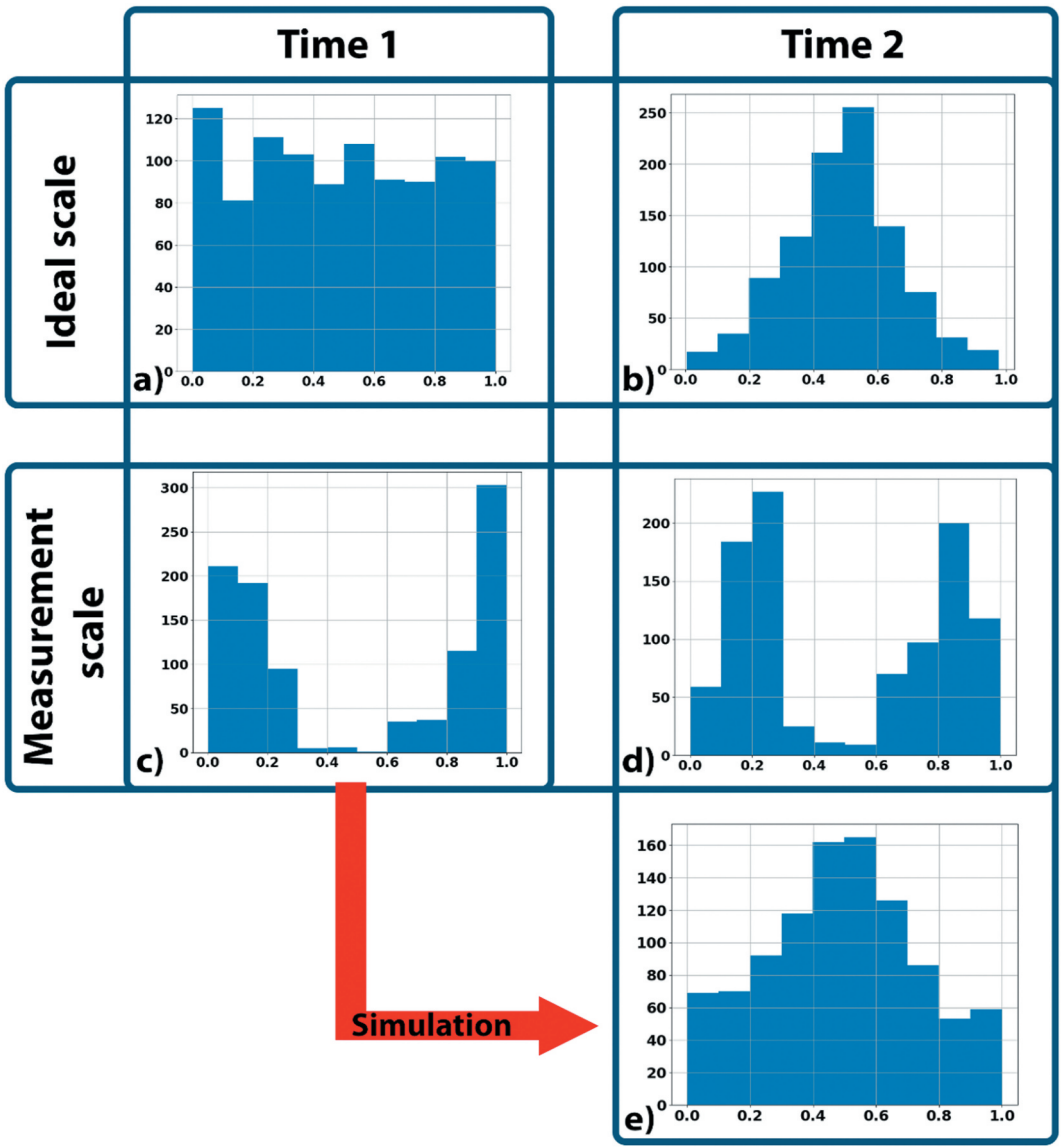

**Figure 8.** Representation of the opinion dynamics for time 1 (a and c) and for time 2 (i.e. after 1,000 time steps) (b, d and e). Figure a and b represent the opinions on the 'ideal scale' while c and d are their transformation to the measurement scale. e) is obtained by seeding the model with c).

This means that poor predictions are not necessarily evidence that a model is poor. Indeed, we may find that the same model can produce much better predictions using another scale, as showed in the previous section. Furthermore, this result is not confined to the act of predicting but it influences also the act of qualitative comparison of model's output.

Similarly, if a model produces high-quality predictions in a specific scenario, it does not mean that it will still produce them in another case. For example, let us suppose that a model M produces extremely good predictions (either quantitative or by qualitative comparison) with a specific measurement of political opinion. If we now try to use the same model on another scale of political opinion, we may end up with completely wrong predictions. Similarly, we do not know how the same model will perform when used with data on trust in science. Indeed, the trust in science scale would be a completely new ordinal scale (as it is based on different questions) with new distortions. Although it is not intuitive, the analysis above demonstrates that the relationship between the measurement scale and the model is fundamental for exploring the model's properties. Unfortunately, the relationship between ABM and scales is, to this point mostly unexplored.

This would be fundamental also for better understanding and developing the process of validation. For example, some authors validate their model by comparing it at the same time against multiple scales (Chattoe-Brown, 2014; Duggins, 2014). In this paper, we discussed how different scales would have different distortions and produce different dynamics in the model (e.g. the Deffuant model producing non-Deffuant-like patterns). Therefore, it is not clear if the fact that a model can reproduce the results of different scales should be considered as a positive result (for example, as the fact that the model can reproduce many different realistic patterns) or as a negative result (for example, as the fact that the model can possibly reproduce every possible pattern).

## Discussion

In this article, we showed that some of the data type used in ABMs are usually affected by psychometric distortions. Distortions are non-linear transformations from one scale to another which can strongly alter the data distribution and, consequently, the model's predictions. This also has important consequences on model's validity, as the quality of the model's predictions strongly depends on the chosen scale. This applies both to the case in which we want to seed data to a model or to the case in which we want to compare model's output to real-world data.

One way to decrease this complexity is similar to the one in physics and it consists in fixing the way the constructs are measured (Halliday et al., 2013). Indeed, in physics if a model makes low-quality predictions, it is considered a poor model, without looking for alternative ways of measuring the physical quantities. In ABMs this would mean that, before testing a model of, let us say, dynamics of political opinions, the community should identify a standard measurement of political opinion. In this way every researcher will test her models against the same reference. However, we are aware that such a solution would be extremely difficult to achieve. Indeed, in many fields new research continually introduces new ways of measuring the same construct.

Another solution would consist in trying to calibrate the values of the ordinal scale. For example, we mentioned that the different scores on ordinal scales should be represented as labels $m_1$, $m_2$, etc. Thus, we could try to guess the values of all these scores that produce the best predictions for a given model. In this way, we could still use models which suppose data from ratio scale, as we would identify the right spacing between the levels (with the measurement model to optimize model fit). In this case, measurement calibration would become an additional model parameter. Furthermore, this will allow to use the same model with data coming from different scales.

Further studies should also focus on how other related properties can affects models' predictions and comparison. For example, both sample size and subjectivity may be studied in the form of measurement error. Notice also that distortions will not affect any model in the same way. Indeed, while for more 'classical' models it is sometimes possible to estimate the impact of distortions for an entire class of models, in ABM, the impact of distortions should be studied for each model separately.

Another interesting possibility comes from developing models directly from experimental data. This will allow to define a precise scale for a model (i.e. the one used for collecting the experimental data). Furthermore, if such a model should then be applied to a different set of data, distortions between the two scales could be studied to re-adapt the model.

In conclusion, while there have been several recent calls for more applications of agent-based models to empirical data (Castellano et al., 2009; Dong et al., 2018; Flache et al., 2017; Valori et al., 2012), we caution that the fit between ABMs and empirical data needs careful consideration. Comparing or seeding ABMs to empirical data is straightforward when measures are on undistorted ratio scales, but it can easily break down when it depends on psychometric measurements. Indeed, a change of measurement scale (e.g. between two different surveys) may strongly affect model's dynamics. Thus, we hope that in the future more work will be dedicated on studying the relationship between different ABMs and data properties, including both distortions and statistical robustness.

## Notes

1. Besides the lack of units, psychometric measurements are often self-reported. This also introduces the problem of subjectivity. For examples, even if two people *report* that they are feeling a pain of 8/10, we still cannot guarantee that they are really *experiencing* the same level of pain. While this is also a very important problem connected to scales, it is beyond the purpose of this paper.
2. Usually, scales are divided in five groups: nominal, ordinal, interval, ratio and cardinal. In this article, we will focus only on ordinal and ratio. Thus, to avoid introducing too many concepts we will not introduce or discuss nominal, cardinal and interval scales. However, the interested reader may find further information in Stevens (Stevens, 1946).
3. While ideal ordinal scales can always distinguish between different levels, real scales have finite sensitivity. Therefore, even if two person have different values of some construct, they may still obtain the same score.
4. Notice that while here we focus on psychometric measurements, the entire analysis can be easily extended to any ordinal measurement.
5. It is important to notice that in the literature there are two similar uses of the term 'ordinal scales.' The most common one (which we also use) follows Stevens (1951) considers Likert-type scales as ordinal. The second use, which is mostly used in mathematical formulation of measurements, would consider Likert-type scales as ordinal scales affected by both noise and limited precision (or granularity). As a detailed discussion is beyond the purpose of this paper, the interested reader may check (Krantz et al., 2006a)
6. Due to its complexity, a full explanation of the Duggins model is not possible in this article and beyond the purpose of this paper. The interested reader may find detailed information in Duggins (2014).

## Acknowledgments

This project has received funding from the European Union's Horizon 2020 research and innovation programme under the Marie Skłodowska-Curie grant agreement No 891347 and from the European Research Council (ERC) under the European Union's Horizon 2020 research and innovation programme (grant agreement No. 802421).

## Disclosure statement

No potential conflict of interest was reported by the author(s).

## Funding

This project has received funding from the European Union's Horizon 2020 research and innovation programme under the Marie Skłodowska-Curie grant agreement No 891347 and from the European Research Council (ERC) under the European Union's Horizon 2020 research and innovation programme (grant agreement No. 802421).

## Notes on contributors

*Dino Carpentras* is a Marie Curie Fellow working on agent-based models applied to vaccine hesitancy. His research focuses on how agent-based models can be used for making predictions for policy making starting from opinion-related data. He also focuses on developing network-based methods for exploring belief networks and opinion spaces.

*Michael Quayle* is a social psychologist holding positions as Associate Professor at the University of Limerick and honorary senior lecturer at the University of KwaZulu-Natal, South Africa. Broadly, his research explores identity, and how identity enactment can impact broader socio-political processes. His work explores these processes across a range of contexts, including gender, race and racism, political speeches, sports, and health education amongst others. Currently he is focussing on understanding how networks of attitudes form a basis for group identity, and how opinion-based identity processes might shape outcomes in social networks.

## ORCID

Dino Carpentras  http://orcid.org/0000-0001-8471-2352
Michael Quayle  http://orcid.org/0000-0002-7497-0566